\documentclass[12pt]{article}
\usepackage[affil-it]{authblk}
\usepackage{url}
\usepackage{graphicx}
\usepackage{fullpage}
\usepackage{amsmath, amssymb}
\usepackage{latexsym}
\usepackage{amsthm}
\usepackage{setspace}
\newtheorem {theo} {\bf Theorem} [section]

\newtheorem {defn} [theo] {\bf Definition}
\newtheorem {rmk}  [theo] {\bf Remark}
\newcommand{\R}{{\mathbb R}}

\newcommand{\Z}{{\mathbb Z}}

\title{Are there Unfoldable Proteins in Dimension Three?}
\author{Rinni Bhansali and Folkert Tangerman\\  \textit{Email: rinni.bhansali@gmail.com and fmtangerman@gmail.com}}
\begin{document}
\date{}
\maketitle

\tableofcontents

\begin{abstract}
In this paper we show the existence of three dimensional rigid, and thus unfoldable, lattice conformations. The structure described here has 461 bonds. We provide a computer assisted proof of its rigidity. The existence of rigid two dimensional structures was shown earlier, see \cite{MS}. This work answers Question 8 in \cite{2D} in the affirmative: rigid (and hence unfoldable from a straight conformation) self avoiding lattice walks exist also in dimension three. The existence of such rigid structures illustrates why protein folding problems are hard in dimension three: it may not possible to fold one conformation into another using a specific set of folding rules. 
\end{abstract}

\newpage
\section{Introduction}

Linear protein structures are modeled as chains of atoms that are connected by bonds and mathematically as self-avoiding walks (SAW) in dimension two or dimension three. Their (two or three) dimensional shape is also often referred to as a conformation of the given protein structure. For a general background information on the vast topic of protein folding we refer to \cite{Dill1, W, LH}.

We consider the following set of problems: given two different three dimensional conformations of a given linear protein structure. Is it possible to fold, by well-defined folding moves, the one into the other while ensuring that intermediate conformations maintain the self avoidance property? In games such as Foldit, see \cite{Foldit}, it is usually a priori given to the player that such a set of folding moves is possible, since the game would otherwise be 'unfair', and it is left up to to the player to determine a set of moves. It is well known that this game becomes increasingly more difficult with increasing length of the protein chain. In a computer simulation of protein folding, for instance for drug discovery, it may similarly be the case that one starts with a conformation that is easy to implement and contains the atoms in the right order, consistent with laboratory data, which then needs to be folded into a physically sensible shape, usually one of much reduced free energy. It is also then the case that with increasing length of the protein chain, to find the minimum energy conformation, or one that is close to a minimum energy conformation becomes rapidly more complicated and appears to be NP-complete in a strict sense, see for instance \cite{NP}.

This paper concentrates on the three dimensional case, with simplified, but rather standard, computational chemistry assumptions on bond-lengths and angles. The set of allowable bond angles is retricted to be multiples of $90^o$, and we assume for simplicity that all bonds have equal length. This allows us to model SAW conformations as three dimensional self avoiding walks along the lattice ${\mathbb Z}^3$. In this paper we exhibit a conformation that is woven so tightly together, as a ball of yarn, that it is completely rigid: every allowed folding move, defined in section 2.2, results in self-intersections. Presented with another conformation of this structure, and these folding rules, the Foldit game would be then be clearly unfair! It will impossible to fold the shape given here into any other shape. As in \cite{2D} we call such conformations also \textbf{unfoldable} since these can not be folded from a straight conformation. Rigid conformations are special: they can not be folded into any other conformations. The set of unfoldable conformations is presumably much richer and is expected to consist of clusters of conformations that can be folded into one another but not into a straight conformation, see \cite{2D} for the analogous discussion in dimension two, and discussed in more detail below.  

\begin{itemize}
\item While our folding moves are perhaps overly restrictive the existence of rigid conformations illustrates why folding problems are intrinsically difficult with increasing length. 
\item
We conjecture that such rigid conformations can be found also when one adopts less restrictive  but still physically realistic ranges of bond angles. In that case then the shape space, the space of conformations, is also disconnected and it would be of interest to find perhaps knot-based methods to describe the connected components of conformation space.
\end{itemize}

\begin{figure}[htb]
\centering
\includegraphics[width=5cm]{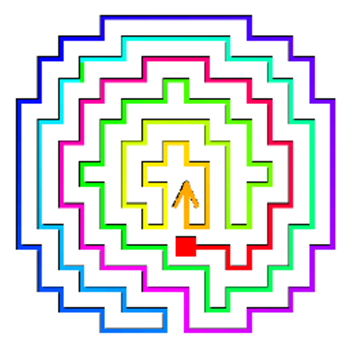}
\caption{The first discovered unfoldable two dimensional self avoiding walk in dimension 2, which was shown to be rigid in \cite{MS}}
\label{2D Figure}
\end{figure}

Our work is inspired by the two dimensional study presented in \cite{2D}, see also \cite{SAWfolding}. These papers \cite{2D,SAWfolding} reveal cluster of conformations that can be folded into one another but not folded from straight conformations. The key ingredient was supplied much earlier in \cite{MS} which showed the existence of rigid two dimensional conformations, as shown in Figure (\ref{2D Figure}). The first rigid two dimensional conformation found had 223 bonds, the shortest found so far still has over a 100 bonds (107) and the shortest one has not yet been found in dimension two.

The problem to \textbf{decide} if a given conformation is rigid tends to be comparatively easy, once the set of allowed moves is clearly described. To find such rigid conformations required a geometrical idea, although it may be possible to find them by 'brute force'. The three dimensional conformation, consists of 461 bonds, presented in this paper is computer constructed and based on the 'ball of yarn' idea, as we will describe in chapter 2. The verification that this structure is rigid is similarly computer assisted. The required Matlab code for its construction is contained in the Appendix.

We point out that a conformation that consists of 461 bonds is still very short on a biological scale: many protein chains consist of thousands of atoms.

\subsection{Acknowledgements} The authors gratefully acknowledge an inspring conversation with Prof. Ken A. Dill, Laufer Center for Physical and Quantitative Biology, Stony Brook University.

\section{Description of the Three Dimensional Structure}

\subsection{Notation for Lattice Conformations}

We consider conformations that consist of chains of atoms with bonds, the connections between atoms, of equal and unit length. We also assume that the bonds are aligned with the primary x, y, and z, coordinate axes. By convention a conformation has a beginning atom at the origin $(0,0,0)$, and the subsequent atoms at lattice locations in $\Z^3$.  Viewed from the beginning atom every bond is oriented with a beginning atom and an end atom and can be in one of six possible orientations. We adopt the following convention +1 (-1) if the bond oriented in the positive (negative) x direction, +2 (-2) if the bond oriented in the positive (negative) y direction, and +3 (-3) if the bond oriented in the positive (negative) x direction. 

With this convention a lattice conformation $c$ of length $n$ is then given as a sequence of $c=(b_1,b_2,...,b_n)$ with each $b_i\in\{+1,-1,+2,-2,+3,-3\}$ for $i=1,...,n$. Such a conformation then consists of $n+1$ atoms, with lattice locations that can then be produced from the sequence $c$, by starting at the beginning atom and traversing the bond sequence, using vector addition.  

\begin{defn}
A lattice conformation $c$ is a self avoiding walk (SAW) if it is properly embedded in $\R^3$, i.e. without self intersections.
\end{defn}  

\begin{rmk}
For lattice conformations to decide if a conformation $c$ is a SAW is straight forward: intersections can only occur at lattice points or at midpoints of bonds where various bonds could intersect. We use the following algorithm to determine if a $c$ is a SAW: generate list of atoms locations by traversing the bonds in $c$, verify if there are any duplicates in this list, by a sort on the integer coordinates of atoms. Do the same for midpoints of bonds. If there are no duplicates in either of these two lists then $c$ is a SAW.
\end{rmk}
 
\subsection{Elementary Folding Moves}

The critical definition used in this paper is that
of an elementary folding move.

Consider a conformation $c_0=(b_1,b_2,...,b_n)$. Consider the i-th bond $b_i$ and decompose $c_0$ as $c_0=c_b b_i c_a$, where $c_b$ consists of the sequence $(b_1,...,b_{i-1})$ and $c_a=(b_{i+1},...,b_n)$. Replace $b_i$ by another bond $b'_i$ and let
$c_1=c_b b'_i c_a$. The {\bf folding motion} we consider is to leave the atoms in $c_b$ fixed, to rotate the bond $b_i$ about its beginning atom to $b'_i$ while translating the atoms in $c_a$ along while maintaining the orientation of the bonds in $c_a$. We denote by $c_t$ $(0\leq t \leq 1)$ the intermediary \textbf{non-lattice} conformation thus formed, a curve in $\R^3$. 

The justification for considering such a folding move to be elementary is that only a single bond changes orientation and we believe this to be a folding move of minimal internal energy change  in the conformation.

Elementary folding moves can be composed over time to produce many more complicated folding moves: pivots, rotations, and complicated wiggly patterns. An important characteristic of these moves is that we assume that during any short time interval at most one such elementary folding move can occur which we believe to be a physically reasonable assumption.
  
The crux of our rigidity algorithm is the following elementary observation.

\begin{theo}
Assume that $c_1$ is obtained from $c_0$ by an elementary folding move. Let $c_t$ $(0 \leq t \leq 1)$  be the intermediary (non-lattice) configuration obtained by the elementary folding move.
Assume that $c_0$ is as SAW.
Then:  $c_1$ is a SAW if and only if for each $(0\leq t \leq 1)$ ea$c_t$ is self avoiding walk in $\R^3$.  
\end{theo}

In other words: $c_1$ is not a SAW if and only if for some $t$, $(0\leq t \leq 1)$, $c_t$ self intersects. 

\begin{defn}
A SAW lattice conformation $c$ is \textbf{rigid} if and only if \textbf{every} $c'$ obtained from $c$ by an elementary folding move is not a SAW.
\end{defn}  

\subsection{Verification of Rigidity Algorithm}

The algorithm we implemented to verify if a given self avoiding conformation $c$ is rigid, and thus unfoldable, is simply to consider all conformations $c'$ obtained from $c$ by an elementary folding move and to verify that $c'$ is \textbf{not} self avoiding.   
This algorithm was implemented in the Matlab code in the Appendix.

\subsection{The Three Dimensional Example}

The conformation created by the authors consists of 461 bonds and is listed below using the orientation conventions described earlier.

Since we are constructing lattice conformation we need to describe the basic segment that we use, and is what we call \textbf{a spiral of length  $n$}. Such a spiral essentially consists of curve of  length $n$ of each side in say the $x$ and $y$ direction which also winds itself in a screw like motion in the $z$ direction for $n$ steps. So, a spiral of length $4$ is rather small, while one of length $8$ is rather large in size. It is these spirals that are the analogs of a layer in lattice approximation of tightly woven ball of yarn. The various spirals are then connected by 'connection' pieces whose only purpose is to connect subsequent spirals.

The detailed conformation and its organization is shown in the listing below, and in Figure (\ref{3D Figure}). The best way to view this structure is run the Matlab code in the Appendix which also visualizes the structure.
   
The beginning atom of the conformation is in the center and the conformation wraps about the center location first in a spiral of length 4, which is then connected to a spiral of length 8. After that, the conformation works its way back to the center via a a spiral of length 6 to guarantee the endpoint is also be buried deep inside, and all open space in the interior is filled up so that both the beginning and the end point of the conformation are fully trapped. The precise bond sequence is the following, where we provide commentary of the left elucidating the overall structure:
\begin{verbatim}
b=[2 1 3 3 3 1 2 2 3 -1 -1 -2 -2 1 1 2 1 -3 -3 -3 

4 Spiral= -1 -2 -2 -2 -2 3 3 3 3 2 2 2 2 -1 -3 -3 -3 -3 -1 -2 -2 
          -2 -2 3 3 3 3 

Connection= 3 3 -1 -1 -1 

8 Spiral= -3 -3 -3 -3 -3 -3 -3 -3 1 1 1 1 1 1 1 1 3 3 3 3 3 3 3 3
           2 -1 -1 -1 -1 -1 -1 -1 -1 -3 -3 -3 -3 -3 -3 -3 -3 1 1
           1 1 1 1 1 1 2 3 3 3 3 3 3 3 3 -1 -1 -1 -1 -1 -1 -1 -1
          -3 -3 -3 -3 -3 -3 -3 -3 2 1 1 1 1 1 1 1 1 3 3 3 3 3 3 3 
           3 -1 -1 -1 -1 -1 -1 -1 -1 2 -3 -3 -3 -3 -3 -3 -3 -3 1 1
           1 1 1 1 1 1 3 3 3 3 3 3 3 3 2 -1 -1 -1 -1 -1 -1 -1 -1 -3 
          -3 -3 -3 -3 -3 -3 -3 1 1 1 1 1 1 1 1 2 3 3 3 3 3 3 3 3
          -1 -1 -1 -1 -1 -1 -1 -1 -3 -3 -3 -3 -3 -3 -3 -3 2 1 1
           1 1 1 1 1 1 3 3 3 3 3 3 3 3 -1 -1 -1 -1 -1 -1 -1 -1 2 
          -3 -3 -3 -3 -3 -3 -3 -3 1 1 1 1 1 1 1 1 3 3 3 3 3 3 3 3

Connection= -1 -1 -1 -1 -1 -1 -1 -3 

6 Spiral= -2 -2 -2 -2 -2 -2 1 -3 -3 -3 -3 -3 -3 2 2 2 2 2 2 3 3 3
           3 3 3 1 -2 -2 -2 -2 -2 -2 -3 -3 -3 -3 2 2 2 2 2 2 1 3 3
           3 3 -2 -2 -2 -2 -2 -2 -3 -3 -3 -3 -3 -3 1 2 2 2 2 2 2 3
           3 3 3 3 3 -2 -2 -2 -2 -2 -2 1 -3 -3 -3 -3 -3 -3 2 2 2 2 
           2 2 3 3 3 3 3 3 1 -2 -2 -2 -2 -2 -2 -3 -3 -3 -3 -3 -3

End=       2 2 2 2 2 2 3 3 3 3 3 -2 -2 -2 -2 -2 -3 -3 -3 -3 2 2 2 2 
           3 3 3 -2 -2 -2 -3 -3 2 3 -1 2 2 3 3 3 -2 -2 -2 -2 -3 -3 
          -3 -3 -1 2 2 1 -3 2 2 3 -1 -1 -3 -1 -2 1 -3 -2 -1 3 -1];
\end{verbatim} 

\begin{figure}[htb]
\centering
\includegraphics[width=15cm]{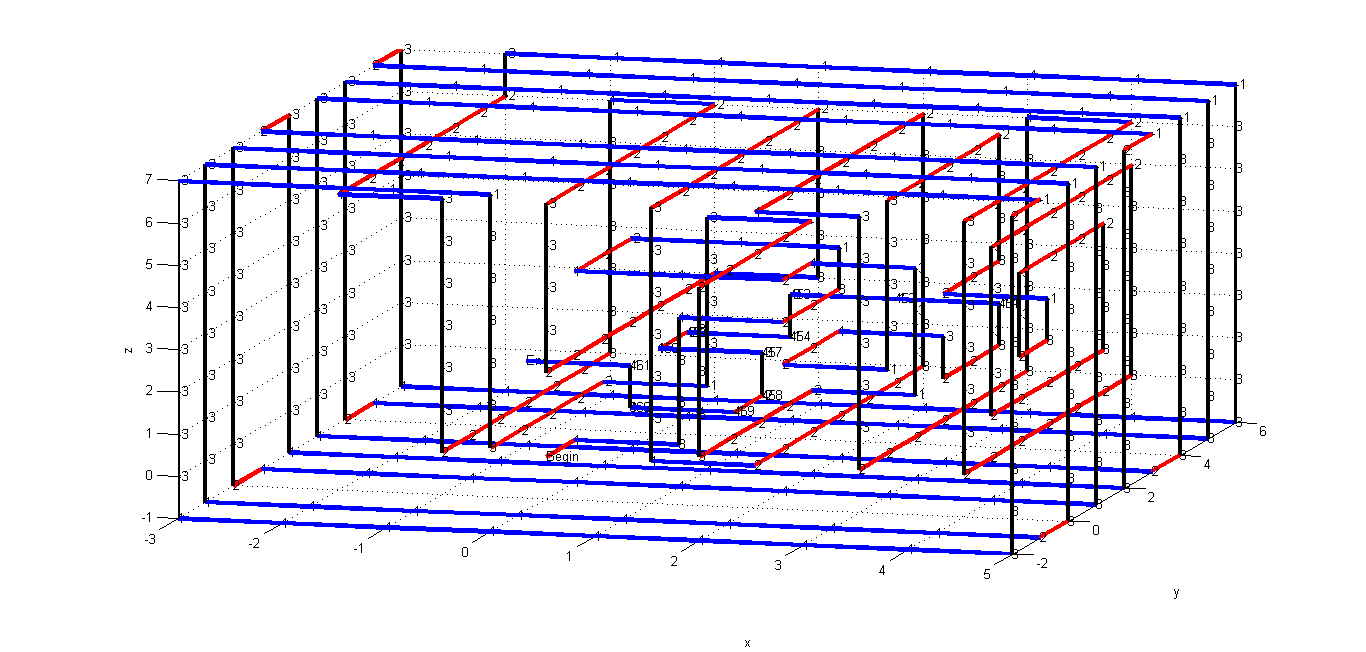}
\caption{Shown is a Matlab rendering of the rigid structure found. The colors indicate the primary bond directions: blue -- x direction, red -- y direction, and black -- z direction. Indicated also are the location of the begin point and the end point of the conformation. The Matlab code attached in the appendix allows one to further rotate and manipulate this structure.}
\label{3D Figure}
\end{figure}

\section{Conclusions}

In this paper we have shown the existence of three dimensional rigid, and hence unfoldable, linear lattice conformations. The conformation demonstrated here has 461 bonds and has the general structure of a tightly woven ball of yarn: its primary weaves are  spiral shaped, and both ends of the conformation are tucked in near the center of the conformation. 

Outstanding questions are:
\begin{itemize}
\item What is the length and shape of the shortest rigid lattice conformation?
\item What does the shape space of lattice conformation look like?
\item Say that two conformations are connected if we can fold the one into the other using elementary moves. What is a typical size of connected components?
\item How to estimate the algorithmic complexity of the question if a given conformation is unfoldable?
\item Do rigid examples exist that also exhibit some kind of mechanical rigidity? We note that the example exhibited here does not seem to be particularly mechanically advantageous.
\end{itemize}

We can of course ask similar questions when considering other sets of three dimensional bond-angles and bond-lengths. 
Note that if we consider the continuous limit with arbitrarily small bond angles then simple rigidity disappears and every conformation is unfoldable. The questions then is what are physically realistic constraints on the folding process.

A final open question is what kind of near term applications rigid of near rigid structures have if they are found in nature or produced in the lab. We can imagine for instance memory devices which use such properties.

\section{Appendix: Matlab Code for Verification.}

\begin{verbatim}
function pfinal1()

%The Unfoldable/Rigid Conformation:
b=[2 1 3 3 3 1 2 2 3 -1 -1 -2 -2 1 1 2 1 -3 -3 -3 -1 -2 -2 -2 -2 ...
    3 3 3 3 2 2 2 2 -1 -3 -3 -3 -3 -1 -2 -2 -2 -2 3 3 3 3 3 3 ...
    -1 -1 -1 -3 -3 -3 -3 -3 -3 -3 -3 1 1 1 1 1 1 1 1 3 3 3 3 3 3 3 3 2 ...
    -1 -1 -1 -1 -1 -1 -1 -1 -3 -3 -3 -3 -3 -3 -3 -3 1 1 1 1 1 1 1 1 2 ...
    3 3 3 3 3 3 3 3 -1 -1 -1 -1 -1 -1 -1 -1 -3 -3 -3 -3 -3 -3 -3 -3 2 ...
    1 1 1 1 1 1 1 1 3 3 3 3 3 3 3 3 -1 -1 -1 -1 -1 -1 -1 -1 2 ...
    -3 -3 -3 -3 -3 -3 -3 -3 1 1 1 1 1 1 1 1 3 3 3 3 3 3 3 3 2 ...
    -1 -1 -1 -1 -1 -1 -1 -1 -3 -3 -3 -3 -3 -3 -3 -3 1 1 1 1 1 1 1 1 ...
    2 3 3 3 3 3 3 3 3 -1 -1 -1 -1 -1 -1 -1 -1 -3 -3 -3 -3 -3 -3 -3 -3 ...
    2 1 1 1 1 1 1 1 1 3 3 3 3 3 3 3 3 -1 -1 -1 -1 -1 -1 -1 -1 2 ...
    -3 -3 -3 -3 -3 -3 -3 -3 1 1 1 1 1 1 1 1 3 3 3 3 3 3 3 3 ...
    -1 -1 -1 -1 -1 -1 -1 -3 -2 -2 -2 -2 -2 -2 1 -3 -3 -3 -3 -3 -3 ...
    2 2 2 2 2 2 3 3 3 3 3 3 1 -2 -2 -2 -2 -2 -2 -3 -3 -3 -3 ...
    2 2 2 2 2 2 1 3 3 3 3 -2 -2 -2 -2 -2 -2 -3 -3 -3 -3 -3 -3 1 ...
    2 2 2 2 2 2 3 3 3 3 3 3 -2 -2 -2 -2 -2 -2 1 -3 -3 -3 -3 -3 -3 ...
    2 2 2 2 2 2 3 3 3 3 3 3 1 -2 -2 -2 -2 -2 -2 -3 -3 -3 -3 -3 -3 ...
    2 2 2 2 2 2 3 3 3 3 3 -2 -2 -2 -2 -2 -3 -3 -3 -3 2 2 2 2 3 3 3 ...
    -2 -2 -2 -3 -3 2 3 -1 2 2 3 3 3 -2 -2 -2 -2 -3 -3 -3 -3 -1 2 2 1 ...
    -3 2 2 3 -1 -1 -3 -1 -2 1 -3 -2 -1 3 -1];

    
c=[];
c.b=b;
c.beginp=[0 0 0];
    
f1=figure;
set(f1','Position',[800 200 500 400]);
hold on
print_bonds=1;
visualize_conformation(c,print_bonds);
drawnow;

if fast_saw(c)~=1
    error('Warning c is not selfavoiding')
end


is_isolated=isrigid(c);
if is_isolated
    disp('The conformation is rigid');
end

return


%% The function aw checks if conformations c1 and c2 intersect
function [val,ac12,mbc12]=aw(c1,c2,to_plot)

if nargin < 3
    to_plot=1;
end
val=0;

ac1=atoms(c1);
all_atoms=1;
ac2=atoms(c2,all_atoms);
ac12=intersect(ac1,ac2,'rows');
if ~isempty(ac12)
    n_atoms=size(ac12,1);
    for k=1:n_atoms
        atom=ac12(k,:);
        if to_plot
            plot3(atom(1),atom(2),atom(3),'pc','Linewidth',4);
        end
    end
end

mbc1=midpoint_of_bond(c1);
mbc2=midpoint_of_bond(c2);
mbc12=intersect(mbc1,mbc2,'rows');
if ~isempty(mbc12)
    n_bonds=size(mbc12,1);
    for k=1:n_bonds
        mbond=mbc12(k,:);
        if to_plot
            plot3(mbond(1),mbond(2),mbond(3),'or','Linewidth',4);
        end
    end
end

if isempty(ac12)&&isempty(mbc12)
    val=1;
    return
end

return

function ac=atoms(c,do_all_atoms)

if nargin < 2
    do_all_atoms=0;
end

ac=[];
beginp=c.beginp;
b=c.b;
ac(1,:)=beginp;
%Says  where random walk begins

if do_all_atoms==0
    n_atoms_to_do=length(b)-1;
else
    n_atoms_to_do=length(b);
end

for k=1:n_atoms_to_do
    direction=b(k);
    if abs(direction)==1 % x direction
        endp=beginp+sign(direction)*[1 0 0];
    elseif abs(direction)==2 % y direction
        endp=beginp+sign(direction)*[0 1 0];
    elseif abs(direction)==3 % z direction
        endp=beginp+sign(direction)*[0 0 1];
    else 
        error('unknown direction')
    end
    ac(k+1,:)=endp; 
    beginp=endp;

end
return

function mbc=midpoint_of_bond(c)

mbc=[];
beginp=c.beginp;
b=c.b;
%Says  where random walk begins
for k=1:length(b)
    direction=b(k);
    if abs(direction)==1 % x direction
        endp=beginp+sign(direction)*[1 0 0];
    elseif abs(direction)==2 % y direction
        endp=beginp+sign(direction)*[0 1 0];
    elseif abs(direction)==3 % z direction
        endp=beginp+sign(direction)*[0 0 1];
    else 
        error('unknown direction')
    end
    mbc(k,:)=(beginp+endp)/2; 
    beginp=endp;

end
return

function visualize_conformation(c,print_bonds,color)

if nargin<2
    print_bonds=1;
end
if nargin<3
    color='b-';
end

beginp=c.beginp;
b=c.b;
xmin=0;xmax=0;
ymin=0;ymax=0;
zmin=0;zmax=0;
%Says  where random walk begins
for k=1:length(b)
    direction=b(k);
    if abs(direction)==1 % x direction
        endp=beginp+sign(direction)*[1 0 0];
    elseif abs(direction)==2 % y direction
        endp=beginp+sign(direction)*[0 1 0];
    elseif abs(direction)==3 % z direction
        endp=beginp+sign(direction)*[0 0 1];
    else 
        error('unknown direction')
    end
    
    if nargin < 3
       if abs(direction)==1
           color='b-';
       elseif abs(direction)==2
           color='r-';
       else
           color='k-';
       end
        
    end
    plot3([beginp(1) endp(1)],[beginp(2) endp(2)],[beginp(3) endp(3)], ...
        color,'Linewidth',3);
        
    if print_bonds
        text_str=[num2str(direction)];
        if (k>1)
            text(beginp(1),beginp(2),beginp(3),text_str);
        end
        text_str=[num2str(k)];
        if (k>450) && (k<470)
            text(beginp(1),beginp(2),beginp(3),text_str);
        end
        if k==1
            text(beginp(1),beginp(2),beginp(3),'Begin');
        end
        if k==length(b)
            text(endp(1),endp(2),endp(3),'End');
        end
    end
    beginp=endp;
    %defines what +-1,+-2,and +-3 means, makes labels for each segment,
    %describing segment number and direction 
end
grid on
xlabel('x')
ylabel('y')
zlabel('z')
set(gcf,'Renderer','zbuffer');
return

%% isrigid : checks if a conformation is rigid
function [is_isolated,k,bond_new,cnew]=isrigid(c)
% Input:
%     conformation c
% Output:
%     is_isolated = 1 if the conformation is isolated
%                  0 otherwise.
% This function checks if the conformation c is isolated, the
% simplest case to prove that a conformation is unfoldable.
% Isolated means that there is no bond that can be rotated by 90 degrees
% and maintain the self avoidance property.
% Such rotations are the smallest motions possible. Wiggles for instance
% are compositions of such rotations.
% 
% Precisely this is what the algorithm does:
% For every bond check if this bond can be rotated
% by 90 degrees.
% Let c_before be the conformation before the bond. During such a rotation
%     c_before does not move.
% Let c_after be the conformation after the bond. During such a rotation
%     c_after is assumed to moves only by a sequence of translations.
%     by the displacement between the endpoint of the initial bond and the
%     end point of the bond_new.
% The algorithm then consists in checking that:
%     1. The rotated bond (bond_new) does not intersect c_before.
%     2. The translation of c_after does not intersect the conformation
%        [c_before bond_new].
%Details to show in actuality the intermediate translations of c_after do 
%not intersect [c_before bond_new] and that we therefore do not need to 
%check separately: 

%Step 1: Let bond_before be the bond in the conformation just before 
%the bond. Let bond_after be the bond in the conformation just after the bond. 
%Give the bond any of the possible orientations bond_new, so that
%bond_new is not in the opposite orientation of bond_before.
%bond_new is not in the opposite orientation of bond_after.
%Verify that the bond can be rotated to bond_new without intersecting
%c_before. This amounts to checking that its end-atom does not
%coincide with an atom on c_before.

%Step 2: Assume that the bond can be rotated to bond_new (from step 1)
%consider the new conformation c_new=[c_before bond_new].
%Translate c_after by the translation that translates its beginning
%point to the end-atom of the new conformation c_new.
%Verify that during this translation no intersections with c_new occur.
%The goal of the remainder is to understand what, if any, intersections with
%c_new can occur during the motion. Since the movement from bond to bond_new
%is small, we expect that the only intersections that occur are
%those that occur at the end of the motion, and not in the 'middle'
%of the motion.
%
%Suppose that bond = 1, then bond_new can be any of 2, -2, 3 and -3.
%Assume that bond_new is equal to 2. 
%
%We need to have a formula for how c_after moves. 
%Assume that beginning atom of the bond is at the origin [0 0 0],
%The end atom is then at the point [1 0 0]. After a rotation by angle
%t (t between 0 and 90 degrees) this end atom moves to
%[cos(t) sin(t) 0]. Its displacement vector is equal to [cos(t)-1 sin(t) 0].
%All the atoms in c_after are displaced by the same amount.
%Since all the atoms are initially on
%the lattice Z^3=ZxZxZ the motion of an atom at location (x,y,z) during
%the translation is equal to [x+cos(t)-1,y+sin(t),z)  with t between
%0 and 90.
%The atoms in c_before are at lattice locations. The point 
%            [x+cos(t)-1,y+sin(t),z] 
%is at lattice locations only when y+sin(t) is an integer. which
%Since y is an integer, sin(t) must then be an integer,
%so t=0 or 90 degrees. So there are no intermediate collisions
%(t strictly between 0 and 90 degrees) of an atom on c_after with
%atoms on c_before.
%We next need to see if such a moving atom can intersect a bond in
%c_before.
%Bonds are line segments parallel to the coordinate axes of the form
%[a b c]+s v, v=+/-[1 0 0], or +/-[0 1 0] or +/-[0 0 1], 0<=s<=1
%with [a b c] on the lattice ZxZxZ. Therefore: on a bond at least
%two of the coordinates are integer. The point [x+cos(t)-1,y+sin(t),z] 
%has two of its coordinates integer only if t=0 or 90.
%Therefore the motion of atoms on c_after does not intersect bonds on 
%c_before, except possibly at t=90 degrees.
%We next need to see when bonds in c_after could possibly intersect
%with c_before.
%       
is_isolated=1;
b=c.b;
n_bonds=length(b);
bond_labels=[-3 -2 -1 1 2 3];
to_plot=0;
for k=1:n_bonds
    bond=b(k);
    %step 1: create the list of possible 90 degree rotations
    new_bonds=setdiff(bond_labels,[bond -bond]);
    %the new bonds are not allowed to be opposite to bond_before
    if k>1         
        new_bonds=setdiff(new_bonds,-b(k-1));
    end
    %and are not allowed to be opposite to the bond_after
    if k<n_bonds
        new_bonds=setdiff(new_bonds,-b(k+1));
    end
    for m=1:length(new_bonds) %now we test all rotations
        bond_new=new_bonds(m);
        if (k>1) && (k < n_bonds)
            c_before=[];
            c_before.beginp=c.beginp;
            c_before.b=c.b(1:k-1);
            c_atoms=atoms(c);
            c_mid=[];
            c_mid.beginp=c_atoms(k,:);
            c_mid.b=bond_new;
            aw_before_and_mid= aw(c_before,c_mid,to_plot);
            if aw_before_and_mid==0 %
               % disp('found intersection c_before and bond_new');
               % disp(['bond number = ' num2str(k)]);
            else %there was no intersection  between c_before and bond_new
                cnew=[];
                cnew.beginp=c_before.beginp;
                cnew.b=[c_before.b bond_new];
                cnew_atoms=atoms(cnew);
                c_after=[];
                c_after.b=c.b(k+1:end);
                c_after.beginp=cnew_atoms(end,:);
                direction =cnew.b(end);
                if abs(direction)==1 %FIXED x-direction
                    c_after.beginp=c_after.beginp+sign(direction)*[1 0 0];
                elseif abs(direction)==2 %FIXED y-direction
                    c_after.beginp=c_after.beginp+sign(direction)*[0 1 0];
                elseif abs(direction)==3 %FIXED z-direction
                    c_after.beginp=c_after.beginp+sign(direction)*[0 0 1];
                end
                aw_mid_and_after=aw(cnew,c_after,to_plot);
                if aw_mid_and_after==0 %
                   % disp('found intersection [c_before bond_new] c_after');
                    %disp(['bond number = ' num2str(k)]);
                else %in this case aw_before_and_mid and aw_mid_and_after both equal 1
                    disp('the new conformation is not isolated')
                    disp(['bond number = ' num2str(k)]);
                    disp(['original bond orientation = ' num2str(bond)]);
                    disp(['rotated bond orientation = ' num2str(bond_new)]);
                    disp(' ');
                    cnew.b=[cnew.b c_after.b];
                    is_isolated=0;
                    %return;
                end
                
            end
        end
        if (k==1) %the bond is the first bond
           
            c_atoms=atoms(c);
            c_mid=[];
            c_mid.beginp=c_atoms(k,:);
            c_mid.b=bond_new;
            
            cnew=[];
            cnew.beginp=c.beginp;
            cnew.b=[bond_new];
            cnew_atoms=atoms(cnew);
            c_after=[];
            c_after.b=c.b(k+1:end);
            c_after.beginp=cnew_atoms(end,:);
            direction =cnew.b(end);
            if abs(direction)==1 %FIXED x-direction
                c_after.beginp=c_after.beginp+sign(direction)*[1 0 0];
            elseif abs(direction)==2 %FIXED y-direction
                c_after.beginp=c_after.beginp+sign(direction)*[0 1 0];
            elseif abs(direction)==3 %FIXED z-direction
                c_after.beginp=c_after.beginp+sign(direction)*[0 0 1];
            end
            aw_mid_and_after=aw(cnew,c_after,to_plot);
            if aw_mid_and_after==0 %
               % disp('found intersection [bond_new] c_after');
               % disp(['bond number = 1']);
            else
                %aw_mid_and_after is equal to 1
                    disp('the new conformation is not isolated')
                    disp(['bond number = ' num2str(k)]);
                    disp(['original bond orientation = ' num2str(bond)]);
                    disp(['rotated bond orientation = ' num2str(bond_new)]);
                    disp(' ');
                    cnew.b=[cnew.b c_after.b];
                    is_isolated=0;
                    %return;
            end
        end
        if (k==n_bonds) %the last one
            c_before=[];
            c_before.beginp=c.beginp;
            c_before.b=c.b(1:k-1);
            c_atoms=atoms(c);
            c_mid=[];
            c_mid.beginp=c_atoms(k,:);
            c_mid.b=bond_new;
            aw_before_and_mid= aw(c_before,c_mid,to_plot);
            if aw_before_and_mid==0 %
                %disp('found intersection c_before and bond_new');
                %disp(['bond number = ' num2str(k)]);
            else
                disp('the new conformation is not isolated')
                disp(['bond number = ' num2str(k)]);
                disp(['original bond orientation = ' num2str(bond)]);
                disp(['rotated bond orientation = ' num2str(bond_new)]);
                disp(' ');
                cnew=[];
                cnew.beginp=c_before.beginp;
                cnew.b=[c_before.b bond_new];
                is_isolated=0;
                %return;
            end
        end
    end
end
cnew=[];          
return;

%% 
function val=fast_saw(c)
%input;
%c; a conformation
%output;
%val=1 if c is self avoiding
%    0 if c is self intersecting.
% usage: val=fast_saw(c)
% The function fast_saw only checks if atoms intersect.

val=1;
all_atoms=1;
ac=atoms(c,all_atoms);
%sortrows this arranges the atoms in lexicographical order
%those that are in the same geometric position will be below 
%each other in the sorted list.
ac=sortrows(ac,[1 2 3]);
n_atoms=size(ac,1);
for k=1:n_atoms-1
    atom=ac(k,:);
    atom_next=ac(k+1,:);
    if atom==atom_next
        plot3(atom(1),atom(2),atom(3),'pc','Linewidth',4);
        val=0;
    end
    
end
return
\end{verbatim}

\end{document}